\documentclass[twocolumn,
amsmath,amssymb]{revtex4}
\usepackage{amssymb}
\usepackage[dvips]{graphicx}
\begin{document}

\textbf{Comment on Multiple Quantum Oscillations in the de
 Haas - van Alphen Spectra of the Underdoped High-Temperature
Superconductor YBa$_2$Cu$_3$O$_{6.5}$}.


Recent observations (\cite{ley,aud} and references therein)  of slow
magneto-oscillations in  some underdoped
 cuprate superconductors  are one of the most striking discoveries in the
 field of superconductivity
since many probes of underdoped cuprates clearly point to a non
Fermi-liquid normal state. Here we argue that a large magnitude and
strong magnetic-field dependence of the residual resistance
invalidate the conventional interpretation of these
magneto-oscillations as a normal-state  dHvA effect.

The normal state in-plane resistivity, $\rho$,  of underdoped
samples with a doping close to YBa$_2$CuO$_{6.5}$ remains almost
flat as a function of temperature above T$_c$ with the magnitude
c.a. $250 \div 300$ $\mu\Omega$cm. It remains flat or even
\emph{increases} at temperatures below T$_c$, when the transition is
shifted by the magnetic field \cite{gant}. Indeed, using the
resistance and the size of a sample, where the oscillations have
been observed, (ref. \cite{ley}, supplement), one obtains $\rho
\thickapprox$ 280 $\mu\Omega$cm in the field $B=60$ Tesla at T=1.5K.
The lowest value of the  Dingle temperature, T$_D=\hbar/(2\pi k_B
\tau)$, is T$_D \geqslant \hbar  \rho e^2 k_F^2/(2\pi^2 m k_B c)$,
where $1/\tau$ is the scattering rate. This expression yields T$_D
\geqslant 25$ K with the Fermi momentum $k_F \approx 1.3$ nm$^{-1}$
and the cyclotron mass $m \approx 1.8m_e$,  found in the framework
of the conventional description of oscillations \cite{ley,aud}($c
\approx 1.17$ nm is the c-axis lattice constant). Such a large
Dingle temperature, several times higher than is found from the
conventional analysis of oscillations \cite{ley,aud}, should make
any dHvA effect undetectable. To be compatible with the observed
oscillation amplitudes the residual resistivity has to be below c.a.
70 $\mu\Omega$cm, which is not the case in YBa$_2$CuO$_{6.5}$ for
any magnetic field above 20 Tesla \cite{ley}, where the sample is
allegedly  in the normal state \cite{aud}. Adding more Fermi-surface
pockets only increases the lowest value of T$_D$.

We suggest that the magnetic oscillations have been observed in the
mixed superconducting state well below the upper critical field,
rather than in the normal state. Our suggestion is supported by a
large magnetoresistance, measured in all experimentally accessible
magnetic fields \cite{ley}, and by an estimate of the upper critical
field, $H_{c2}= (\phi_0 k_B \text{T}_c m)^2/\hbar^4 F \gtrsim 100$
Tesla (here $\phi_0$ is the flux quantum,  T$_c=50$K, $m=2m_e$, and
the oscillation frequency $F=500$ Tesla). In fact this mean-field
expression grossly \emph{underestimates} $H_{c2}$, as follows from a
number of measurements of underdoped samples with a sharp in-plane
resistive transition in the magnetic field \cite{gant}. We propose
that the oscillations are most likely caused by a quantum
interference of vortex and crystal lattice modulations of the
superconducting order parameter \cite{ale}.
\begin{figure}
\begin{center}
\includegraphics[angle=-90
,width=0.40\textwidth]{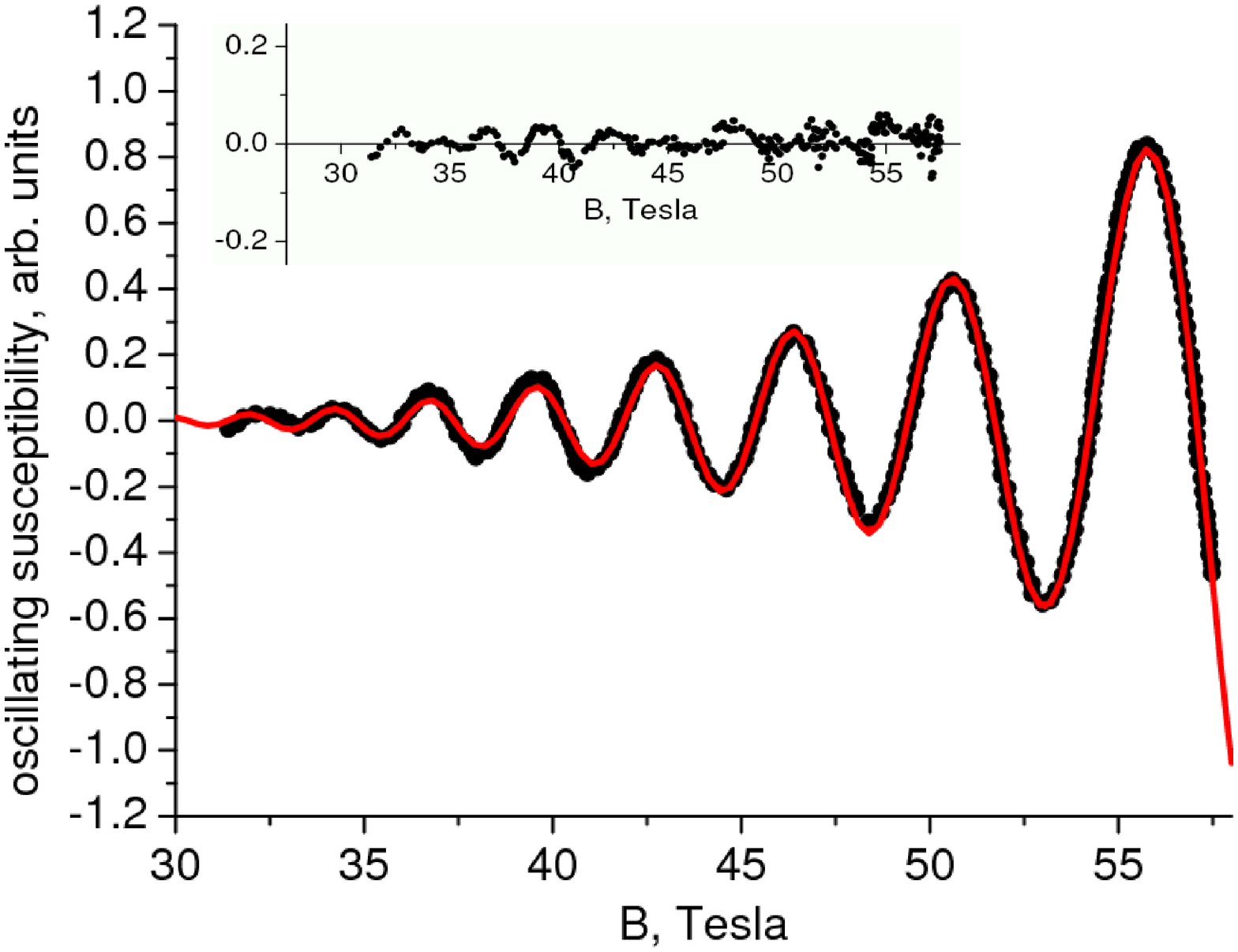} \caption{(Color online) Oscillating
part of the vortex lattice  susceptibility (solid line) compared
with the experimental oscillations \cite{aud} in
YBa$_2$Cu$_3$O$_{6.51}$ (symbols).}
\end{center}
\end{figure}
An excellent fit
of the oscillations in Fig.1 can be obtained with a simple
theoretical expression describing an oscillating part of the
magnetic response of the vortex lattice, $\Delta \chi (B) =
B^{-2}\sum_{i=a,b} A_i e ^{-b_i/B} \cos(\sqrt{B_i/B})$.  As shown by
the residue (data minus fit, Fig.1 inset) the quality of our fit  is
comparable with the conventional fit \cite{aud} but the number of
fitting parameters is several times smaller. Unlike
 the conventional description \cite{aud}, where four oscillation
frequencies emerge from the blue, our characteristic fields fitting
the experiment, $B_a=1.09593\times 10^6$ Tesla and
$B_b=1.02776\times 10^6$ Tesla, are remarkably close to the
theoretical   fields $B_a=8\pi^3 \hbar/ea^2 \approx 1.118 \times
10^6$ Tesla and $B_b=8\pi^3 \hbar/eb^2\approx 1.079 \times 10^6$
Tesla  defined by the crystal lattice in-plane constants, $a\approx
0.382$nm and $b\approx 0.389$nm, respectively \cite{ale}. If the
vortex lattice contains triangular and  square domains, the number
($i$) of characteristic fields contributing to $\Delta \chi (B)$
could be larger than two. As a result we conclude that C. Proust and
colleagues \cite{ley,aud} have observed a novel quantum interference
phenomenon in the mixed state of underdoped cuprate superconductors
proposed by one of us (ASA) \cite{ale}, rather than the conventional
dHvA effect.

We gratefully acknowledge valuable discussions with Nigel Hussey and
Viktor Kabanov and financial support of one of us (IOT) from the
SEPON project (EC VII FP, contract number ERC-2008-AdG-227457).

 A. S. Alexandrov$^1$ and I. O. Thomas$^2$

$^1$Department of Physics, Loughborough University, Loughborough
LE11 3TU, United Kingdom.

$^2$Institute per i Processi Chimici Fisici, Consiglio Nazionale
della Richerche, Via G. Morruzzi 1, 156124 Pisa, Italy.

\end{document}